\documentclass[11pt]{article}

\usepackage{graphicx}

\title{Simplified Variational Principles for Barotropic Magnetohydrodynamics}
\author{Asher Yahalom$^{a,b}$ and Donald Lynden-Bell$^{a,c}$ \\
$^a$ Institute of Astronomy, University of Cambridge\\
Madingley Road, Cambridge CB3 0HA, United Kingdom\\
$^b$ Ariel University Center of Samaria, Ariel 40700, Israel\\
$^c$ Clare College, University of Cambridge, Cambridge, United Kingdom\\
e-mail: dlb@ast.cam.ac.uk; asya@yosh.ac.il; }

\begin{document}
\maketitle

\newcommand{\beq} {\begin{equation}}
\newcommand{\enq} {\end{equation}}
\newcommand{\ber} {\begin {eqnarray}}
\newcommand{\enr} {\end {eqnarray}}
\newcommand{\eq} {equation}
\newcommand{\eqn} {equation }
\newcommand{\eqs} {equations }
\newcommand{\ens} {equations}
\newcommand{\mn}  {{\mu \nu}}
\newcommand {\er}[1] {equation (\ref{#1}) }
\newcommand {\ern}[1] {equation (\ref{#1})}
\newcommand {\ers}[1] {equations (\ref{#1})}
\newcommand {\Er}[1] {Equation (\ref{#1}) }

\begin {abstract}

Variational principles for magnetohydrodynamics were introduced by
previous authors both in Lagrangian and Eulerian form. In this
paper we introduce simpler Eulerian variational principles from which all the relevant
 equations of barotropic magnetohydrodynamics can be derived.
 The variational principle is given in terms of six independent functions
for non-stationary barotropic flows and three independent functions for
stationary barotropic flows. This is less then the seven variables which
appear in the standard equations of barotropic magnetohydrodynamics which are
the magnetic field $\vec B$ the velocity field $\vec v$ and the
density $\rho$.

The equations obtained for non-stationary barotropic magnetohydrodynamics resemble the equations of
Frenkel, Levich \& Stilman \cite{FLS}. The connection between the Hamiltonian formalism
introduced in \cite{FLS} and the present Lagrangian formalism (with Eulerian variables)
will be discussed.

Finally the relations between barotropic magnetohydrodynamics topological constants and the
functions of the present formalism will be elucidated.

\end {abstract}
\\
\\
PACS number(s): 47.65.+a

\section {Introduction}

Variational principles for magnetohydrodynamics were introduced by
previous authors both in Lagrangian and Eulerian form. Sturrock
\cite{Sturrock} has discussed in his book a Lagrangian variational
formalism for magnetohydrodynamics. Vladimirov and Moffatt
\cite{Moffatt} in a series of papers have discussed an Eulerian
variational principle for incompressible magnetohydrodynamics.
However, their variational principle contained three more
functions in addition to the seven variables which appear in the
standard equations of magnetohydrodynamics which are the magnetic
field $\vec B$ the velocity field $\vec v$ and the density $\rho$.
Kats \cite{Kats} has generalized Moffatt's work for compressible
non barotropic flows but without reducing the number of functions
and the computational load. Moreover, Kats has shown that the
variables he suggested can be utilized to describe the motion of
arbitrary discontinuity surfaces \cite{Kats3,Kats4}. Sakurai
\cite{Sakurai} has introduced a two function Eulerian variational
principle for force-free magnetohydrodynamics and used it as a
basis of a numerical scheme, his method is discussed in a book by
Sturrock \cite{Sturrock}. A method of solving the equations for
those two variables was introduced by Yang, Sturrock \& Antiochos
\cite{Yang}. In this work we will combine the Lagrangian of
Sturrock \cite{Sturrock} with the Lagrangian of Sakurai
\cite{Sakurai} to obtain an {\bf Eulerian} Lagrangian principle
which will depend on only six functions. The variational
derivative of this Lagrangian will give us all the equations
needed to describe barotropic magnetohydrodynamics without any
additional constraints. The equations obtained resemble the
equations of Frenkel, Levich \& Stilman \cite{FLS} (see also
\cite{Zakharov}). The connection between the Hamiltonian formalism
introduced in \cite{FLS} and the present Lagrangian formalism
(with Eulerian variables) will be discussed. Furthermore, we will
show that for stationary flows three functions will suffice in
order to describe a Lagrangian principle for barotropic
magnetohydrodynamics. The non-singlevaluedness of the functions
appearing in the reduced representation of barotropic
magnetohydrodynamics will be discussed in particular with
connection to the topological invariants of magnetic and cross
helicities. It will be shown how the conservation of cross
helicity can be easily generated using the Noether theorem and the
variables introduced in this paper.

Due to space limitations this paper is concerned only with barotropic magnetohydrodynamics.
Variational principles of non barotropic magnetohydrodynamics can be found in the work
of Bekenstein \& Oron \cite{Bekenstien} in terms of 15 functions and V.A. Kats \cite{Kats}
in terms of 20 functions. The authors of this paper suspect that this number can be somewhat reduced.
Moreover, A. V. Kats  in a remarkable paper \cite{Kats2} (section IV,E) has shown that there is a
large symmetry group (gauge freedom)  associated with the choice of those functions,
this implies that the number of degrees of freedom can be reduced.

We anticipate applications of this study both to linear and non-linear stability analysis of known
 barotropic magnetohydrodynamic configurations \cite{VMI,AHH}
 and for designing efficient numerical schemes for integrating
the equations of fluid dynamics and magnetohydrodynamics \cite{Yahalom,YahalomPinhasi,
YahPinhasKop,OphirYahPinhasKop}.

The plan of this paper is as follows: first we introduce the
standard notations and equations of barotropic magnetohydrodynamics. Next we review the Lagrangian
variational principle of barotropic magnetohydrodynamics. This is followed by a review of the Eulerian
variational principles of force-free magnetohydrodynamics. After those introductory sections
we will present the six function Eulerian variational principles for non-stationary
magnetohydrodynamics. A derivation of the canonical momenta of the generalized coordinates
appearing in the Lagrangian allows us to derive the system's Hamiltonian which  resembles
the Hamiltonian introduced by Frenkel, Levich \& Stilman \cite{FLS}.
This is followed by the derivation  of a variational principle
for stationary magnetohydrodynamics. The discussion related to the magnetohydrodynamic
topological constants concludes our paper.

\section{The standard formulation of barotropic magnetohydrodynamics}

\subsection{Basic equations}

The standard set of \eqs solved for barotropic magnetohydrodynamics are given below:
\beq
\frac{\partial{\vec B}}{\partial t} = \vec \nabla \times (\vec v \times \vec B),
\label{Beq}
\enq
\beq
\vec \nabla \cdot \vec B =0,
\label{Bcon}
\enq
\beq
\frac{\partial{\rho}}{\partial t} + \vec \nabla \cdot (\rho \vec v ) = 0,
\label{masscon}
\enq
\beq
\rho \frac{d \vec v}{d t}=
\rho (\frac{\partial \vec v}{\partial t}+(\vec v \cdot \vec \nabla)\vec v)  = -\vec \nabla p (\rho) +
\frac{(\vec \nabla \times \vec B) \times \vec B}{4 \pi}.
\label{Euler}
\enq
The following notations are utilized: $\frac{\partial}{\partial t}$ is the temporal derivative,
$\frac{d}{d t}$ is the temporal material derivative and $\vec \nabla$ has its
standard meaning in vector calculus. $\vec B$ is the magnetic field vector, $\vec v$ is the
velocity field vector and $\rho$ is the fluid density. Finally $p (\rho)$ is the pressure which
we assume depends on the density alone (barotropic case). The justification for those \eqs
and the conditions under which they apply can be
found in standard books on magnetohydrodynamics (see for example \cite{Sturrock}). \Er{Beq}describes the
fact that the magnetic field lines are moving with the fluid elements ("frozen" magnetic field lines),
 \ern{Bcon} describes the fact that
the magnetic field is solenoidal, \ern{masscon} describes the conservation of mass and \ern{Euler}
is the Euler equation for a fluid in which both pressure
and Lorentz magnetic forces apply. The term:
\beq
\vec J =\frac{\vec \nabla \times \vec B}{4 \pi},
\label{J}
\enq
is the electric current density which is not connected to any mass flow.
The number of independent variables for which one needs to solve is seven
($\vec v,\vec B,\rho$) and the number of \eqs (\ref{Beq},\ref{masscon},\ref{Euler}) is also seven.
Notice that \ern{Bcon} is a condition on the initial $\vec B$ field and is satisfied automatically for
any other time due to \ern{Beq}. Also notice that $p (\rho)$ is not a variable rather it is a given
function of $\rho$.

\subsection{Lagrangian variational principle of magnetohydrodynamics}

A Lagrangian variational principle for barotropic magnetohydrodynamics has been discussed by a number of authors
(see for example \cite{Sturrock}) and an outline of this approach is given below.
Consider the action:
\ber
A & \equiv & \int {\cal L} d^3 x dt,
\nonumber \\
{\cal L} & \equiv & \rho (\frac{1}{2} \vec v^2 - \varepsilon (\rho)) - \frac{\vec B^2}{8 \pi},
\label{Lagaction}
\enr
in which $\varepsilon (\rho)$ is the specific internal energy.
A variation in any quantity $F$ for a fixed position $\vec r$ is denoted as $\delta F$ hence:
\ber
\delta A & = & \int \delta {\cal L} d^3 x dt,
\nonumber \\
\delta {\cal L} & = &  \delta \rho (\frac{1}{2} \vec v^2 - w (\rho))
+\rho \vec v \cdot \delta  \vec v - \frac{\vec B \cdot \delta \vec B}{4 \pi},
\label{delLagaction}
\enr
in which $w=\frac{\partial (\varepsilon \rho)}{\partial \rho}$ is the specific enthalpy.

A change in a position of a fluid element located at a position $\vec r$ at
time $t$ is given by $\vec \xi (\vec r,t)$. A mass conserving variation of $\rho$
takes the form:
\beq
\delta \rho = - \vec \nabla \cdot (\rho \vec \xi)
\label{delrho}
\enq
and a magnetic flux conserving variation takes the form:
\beq
\delta \vec B = \vec \nabla \times (\vec \xi \times \vec B).
\label{delB}
\enq
A change involving both a local variation coupled with a change of element position of the
quantity $F$ is given by:
\beq
\Delta F = \delta F + (\vec \xi \cdot \vec \nabla) F,
\label{delF}
\enq
hence
\beq
\Delta \vec v = \delta \vec v + (\vec \xi \cdot \vec \nabla) \vec v.
\enq
However, since:
\beq
\Delta \vec v = \Delta \frac{d \vec r}{d t} = \frac{d \Delta \vec r}{d t} =
\frac{d \vec \xi}{d t}.
\enq
We obtain:
\beq
\delta \vec v = \frac{d \vec \xi}{d t} - (\vec \xi \cdot \vec \nabla) \vec v =
\frac{\partial \vec \xi}{\partial t}+(\vec v \cdot \vec \nabla)\vec \xi -
(\vec \xi \cdot \vec \nabla) \vec v.
\label{delv}
\enq
Introducing the result of \eqs (\ref{delrho},\ref{delB},\ref{delv})
into \er{delLagaction} and integrating
by parts we arrive at the result:
\ber
\delta A & = & \int d^3 x \rho \vec v \cdot \vec \xi|^{t_1}_{t_0}
\nonumber \\
& + &
\int dt \{ \oint d \vec S \cdot [-\rho \vec \xi (\frac{1}{2} \vec v^2 - w (\rho))
+ \rho \vec v (\vec v \cdot \vec \xi) + \frac{1}{4 \pi} \vec B \times (\vec \xi \times \vec B)]
\nonumber \\
& + & \int d^3 x  \vec \xi \cdot [-\rho \vec \nabla w - \frac{\partial (\rho \vec v)}{\partial t }
- \frac{\partial (\rho \vec v v_k)}{\partial x_k} - \frac{1}{4 \pi} \vec B \times  (\vec \nabla \times \vec B)
 ]\},
\label{delLagaction2}
\enr
in which a summation convention is assumed. Taking into account the continuity \ern{masscon} we obtain:
\ber
\delta A & = & \int d^3 x \rho \vec v \cdot \vec \xi|^{t_1}_{t_0}
\nonumber \\
& + &
\int dt \{ \oint d \vec S \cdot [-\rho \vec \xi (\frac{1}{2} \vec v^2 - w (\rho))
+ \rho \vec v (\vec v \cdot \vec \xi) + \frac{1}{4 \pi} \vec B \times (\vec \xi \times \vec B)]
\nonumber \\
& + & \int d^3 x  \vec \xi \cdot [-\rho \vec \nabla w - \rho \frac{\partial \vec v}{\partial t}
- \rho (\vec v \cdot \vec \nabla) \vec v  - \frac{1}{4 \pi} \vec B \times  (\vec \nabla \times \vec B ) ]\},
\label{delLagaction3}
\enr
hence we see that if $\delta A = 0$ for a $\vec \xi$ vanishing at the initial and final times and
on the surface of the domain but otherwise arbitrary then Euler's \ern{Euler} is satisfied
(taking into account that in the barotropic case $\vec \nabla w = \frac{\vec \nabla p}{\rho}$).

Although the variational principle does give us the correct dynamical equation for an arbitrary
$\vec \xi$, it has the following deficiencies:
\begin{enumerate}
    \item Although $\vec \xi$ is quite arbitrary the variations of $\delta \rho$ and
    $\delta \vec B$ are not. They are defined by the conditions given in \ern{delrho} and
    \ern{delB}. This property is not useful for numerical schemes since $\vec \xi$ must be a small
    quantity.
    \item Only \ern{Euler} is derived from the variational principle the other equations that are needed:
    \ern{Beq}, \ern{Bcon} and \ern{masscon} are separate assumptions. Moreover \ern{masscon} is
    needed in order to derive Euler's \ern{Euler} from the variational principle.
    All this makes the variational principle less useful.
 \end{enumerate}

 What is desired is a variational principle from which all equations of motion can be derived and
 for which no assumptions on the variations are needed this will be discussed in the following
 sections.

 The reader should also notice two recent interesting papers by Reinhard Prix \cite{Prix1,Prix2}
 which discuss the implications of a  time shift $\tau$ in addition to
 the spatial shift $\vec \xi (\vec r,t)$ and also considers the case of multi fluid magnetohydrodynamics.

\section{Sakurai's variational principle of force-free magnetohydrodynamics}

Force-free magnetohydrodynamics is concerned with the case that both the pressure and
inertial terms in Euler \ers{Euler} are physically insignificant. Hence the Euler equations
can be written in the form:
\beq
\frac{(\vec \nabla \times \vec B) \times \vec B}{4 \pi} = \vec J \times \vec B = 0.
\label{Eulerforce-free}
\enq
In order to describe force-free fields Sakurai \cite{Sakurai} has proposed to represent
 the magnetic field in the following form:
\beq
\vec B = \vec \nabla \chi \times \vec \nabla \eta.
\label{Bsakurai}
\enq
Hence $\vec B$ is orthogonal both to $\vec \nabla \chi$ and $\vec \nabla \eta$.
A similar representation was suggested by Dungey \cite[p. 31]{Dungey} but not in the
context of variational analysis. Frenkel, Levich \& Stilman \cite{FLS} has discussed
the validity of the above representation and have concluded that for a vector field
in the Euclidean space ${\cal R}^3$ the above presentation does always exist locally
but not always globally.  Also Notice that either $\chi$ or $\eta$ (or both) can be
{\bf non single valued functions} (see \cite{FLS} equation 20).

Both $\chi$ and $\eta$ are Clebsch type comoving scalar fields satisfying the
equations:
\beq
\frac{d \chi}{dt} = 0, \qquad \frac{d \eta}{dt} = 0.
\label{comoving}
\enq
It can be easily shown that provided that $\vec B$ is in the form given in \ern{Bsakurai},
and \ern{comoving} is satisfied, then both \ern{Beq} and \ern{Bcon} are satisfied.
Since according to \ern{Eulerforce-free} both $\vec \nabla \times \vec B$ and $\vec B$
are parallel it follows that \ern{Eulerforce-free} can be written as:
\beq
\vec J \cdot \vec \nabla \chi = 0, \qquad \vec J \cdot \vec \nabla \eta = 0.
\label{Eulerforce-free2}
\enq
Sakurai \cite{Sakurai} has introduced an action principle from which \er{Eulerforce-free2} can be derived:
\ber
A_S & \equiv & \int {\cal L_S} d^3 x dt,
\nonumber \\
{\cal L}_S & \equiv & \frac{\vec B^2}{8 \pi} = \frac{(\vec \nabla \chi \times \vec \nabla \eta)^2}{8 \pi}.
\label{LagactionSak}
\enr
Taking the variation of \er{LagactionSak} we obtain:
\ber
\delta A_S & = & \int \delta {\cal L_S} d^3 x dt,
\nonumber \\
\delta {\cal L}_S & = & \frac{\vec B}{4 \pi} \cdot
(\vec \nabla \delta \chi \times \vec \nabla \eta + \vec \nabla \chi \times \vec \nabla \delta \eta).
\label{delLagactionSak}
\enr
Integrating by parts and using the theorem of Gauss one obtains the result:
\ber
\delta A_S & = & \oint d \vec S \cdot [(\delta \chi \vec \nabla \eta -
 \delta \eta \vec \nabla \chi) \times \frac{\vec B}{4 \pi}]+
 \int d \vec \Sigma \cdot [([\delta \chi] \vec \nabla \eta -
 [\delta \eta] \vec \nabla \chi) \times \frac{\vec B}{4 \pi}]
\nonumber \\
 & + &   \int d^3 x [\delta \chi(\vec \nabla \eta \cdot \vec J) - \delta \eta (\vec \nabla \chi \cdot \vec J )]
\label{delLagactionSak2} \enr
in which $\int d \vec \Sigma$ represents an integral along the cut and $[\delta f]$ represents the discontinuity
of the variations of non single valued functions.
We shall show later that $\chi$ can be defined as a single valued function
 while $\eta$ can be either single valued or non single valued.
 Hence if $\delta A_S=0$ for arbitrary variation $\delta \chi,\delta \eta$ that vanish on the
boundary of the domain (including the cut) one recovers the force-free Euler
\ers{Eulerforce-free2}.

Although this approach is better than the one described in \er{Lagaction} in the previous section
in the sense that the form of the variations $\delta \chi,\delta \eta$ is not constrained, it has
some limitations as follows:
\begin{enumerate}
    \item Sakurai's approach by design is only meant to deal with force-free magnetohydrodynamics;
    for more general magnetohydrodynamics it is not adequate.
    \item Sakurai's action given by \ern{LagactionSak} contains all the relevant physical \eqs
     only if the configuration is
    static ($\vec v =0$). If the configuration is not static one needs to supply an additional
    two \ers{comoving} to the variational principle.
   \end{enumerate}

\section{Simplified variational principle of non-stationary barotropic magnetohydrodynamics}

In the following section we will combine the approaches described
in the previous sections in order to obtain a variational
principle of non-stationary barotropic magnetohydrodynamics such that all the
relevant barotropic magnetohydrodynamic equations can be derived from using
unconstrained variations. The approach is based on a
method first introduced by Seliger \& Whitham \cite{Seliger}.
Consider the action:
\ber A & \equiv & \int {\cal L} d^3 x dt,
\nonumber \\
{\cal L} & \equiv & {\cal L}_1 + {\cal L}_2,
\nonumber \\
{\cal L}_1 & \equiv & \rho (\frac{1}{2} \vec v^2 - \varepsilon (\rho)) +  \frac{\vec B^2}{8 \pi},
\nonumber \\
{\cal L}_2 & \equiv & \nu [\frac{\partial{\rho}}{\partial t} + \vec \nabla \cdot (\rho \vec v )]
- \rho \alpha \frac{d \chi}{dt} - \rho \beta \frac{d \eta}{dt} -
\frac{\vec B}{4 \pi} \cdot (\vec \nabla \chi \times \vec \nabla \eta).
\label{Lagactionsimp}
\enr
Obviously $\nu,\alpha,\beta$ are Lagrange multipliers which were inserted in such a
way that the variational principle will yield the following \ens:
\ber
& & \frac{\partial{\rho}}{\partial t} + \vec \nabla \cdot (\rho \vec v ) = 0,
\nonumber \\
& & \rho \frac{d \chi}{dt} = 0,
\nonumber \\
& & \rho \frac{d \eta}{dt} = 0.
\label{lagmul}
\enr
It {\bf is not} assumed that $\nu,\alpha,\beta$  are single valued.
Provided $\rho$ is not null those are just the continuity \ern{masscon} and
the conditions that Sakurai's functions are comoving as in \ern{comoving}.
Taking the variational derivative with respect to $\vec B$ we see that
\beq
\vec B = \hat {\vec B} \equiv \vec \nabla \chi \times \vec \nabla \eta.
\label{Bsakurai2}
\enq
Hence $\vec B$ is in Sakurai's form and satisfies \ern{Bcon}.
By virtue of \ers{lagmul} we see that $\vec B$ must also satisfy \ern{Beq}.
For the time being we have showed that all the equations of barotropic magnetohydrodynamics can be obtained
from the above variational principle except Euler's equations. We will now
show that Euler's equations can be derived from the above variational principle
as well. Let us take an arbitrary variational derivative of the above action with
respect to $\vec v$, this will result in:
\beq
\delta_{\vec v} A = \int d^3 x dt \rho \delta \vec v \cdot
[\vec v - \vec \nabla \nu - \alpha \vec \nabla \chi - \beta \vec \nabla \eta]
+ \oint d \vec S \cdot \delta \vec v \rho \nu+ \int d \vec \Sigma \cdot \delta \vec v \rho [\nu].
\label{delActionv}
\enq
The integral $\oint d \vec S \cdot \delta \vec v \rho \nu$ vanishes in many physical scenarios.
In the case of astrophysical flows this integral will vanish since $\rho=0$ on the flow
boundary, in the case of a fluid contained
in a vessel no flux boundary conditions $\delta \vec v \cdot \hat n =0$ are induced
($\hat n$ is a unit vector normal to the boundary). The surface integral $\int d \vec \Sigma$
 on the cut of $\nu$ vanishes in the case that the flow has zero cross helicity (see section \ref{Top})
 since in this case $\nu$ is single valued and $[\nu]=0$ .
In the case that that the flow has non zero cross helicity, $\nu$ is not single valued (see section \ref{Top}),
in this case only a Kutta type velocity perturbation \cite{YahPinhasKop} in which
the velocity perturbation is parallel to the cut will cause the cut integral to vanish.

Provided that the surface integrals do vanish and that $\delta_{\vec v} A =0$ for an arbitrary
velocity perturbation we see that $\vec v$ must have the following form:
\beq
\vec v = \hat {\vec v} \equiv \vec \nabla \nu + \alpha \vec \nabla \chi + \beta \vec \nabla \eta.
\label{vform}
\enq
Let us now take the variational derivative with respect to the density $\rho$ we obtain:
\ber
\delta_{\rho} A & = & \int d^3 x dt \delta \rho
[\frac{1}{2} \vec v^2 - w  - \frac{\partial{\nu}}{\partial t} -  \vec v \cdot \vec \nabla \nu]
\nonumber \\
 & + & \oint d \vec S \cdot \vec v \delta \rho  \nu +
  \int d \vec \Sigma \cdot \vec v \delta \rho  [\nu] + \int d^3 x \nu \delta \rho |^{t_1}_{t_0}.
\label{delActionrho}
\enr
Hence provided that $\oint d \vec S \cdot \vec v \delta \rho  \nu$ vanishes on the boundary of the domain
and $ \int d \vec \Sigma \cdot \vec v \delta \rho  [\nu]$ vanishes on the cut of $\nu$
in the case that $\nu$ is not single valued\footnote{Which entails either a Kutta type
condition for the velocity or a vanishing density perturbation on the cut.}
and in initial and final times the following \eqn must be satisfied:
\beq
\frac{d \nu}{d t} = \frac{1}{2} \vec v^2 - w.
\label{nueq}
\enq
Finally we have to calculate the variation with respect to both $\chi$ and $\eta$
this will lead us to the following results:
\ber
\delta_{\chi} A & = & \int d^3 x dt \delta \chi
[\frac{\partial{(\rho \alpha)}}{\partial t} +  \vec \nabla \cdot (\rho \alpha \vec v)-
\vec \nabla \eta \cdot \vec J]
+ \oint d \vec S \cdot [\frac{\vec B}{4 \pi} \times \vec \nabla \eta - \vec v \rho \alpha]\delta \chi
 \nonumber \\
 & + & \int d \vec \Sigma \cdot [\frac{\vec B}{4 \pi} \times \vec \nabla \eta - \vec v \rho \alpha][\delta \chi]
 - \int d^3 x \rho \alpha \delta \chi |^{t_1}_{t_0},
\label{delActionchi}
\enr
\ber
\delta_{\eta} A & = & \int d^3 x dt \delta \eta
[\frac{\partial{(\rho \beta)}}{\partial t} +  \vec \nabla \cdot (\rho \beta \vec v)+
\vec \nabla \chi \cdot \vec J]
+ \oint d \vec S \cdot [\vec \nabla \chi \times \frac{\vec B}{4 \pi} - \vec v \rho \beta]\delta \eta
\nonumber \\
 & + & \int d \vec \Sigma \cdot [\vec \nabla \chi \times \frac{\vec B}{4 \pi} - \vec v \rho \beta][\delta \eta]
 - \int d^3 x \rho \beta \delta \eta |^{t_1}_{t_0}.
\label{delActioneta}
\enr
Provided that the correct temporal and boundary conditions are met with
respect to the variations $\delta \chi$ and $\delta \eta$ on the domain boundary and
on the cuts in the case that some (or all) of the relevant functions are non single valued.
we obtain the
following set of equations:
\beq
\frac{d \alpha}{dt} = \frac{\vec \nabla \eta \cdot \vec J}{\rho}, \qquad
\frac{d \beta}{dt} = -\frac{\vec \nabla \chi \cdot \vec J}{\rho},
\label{albetaeq}
\enq
in which the continuity \ern{masscon} was taken into account.

\subsection{Euler's equations}

We shall now show that a velocity field given by \ern{vform}, such that the
\eqs for $\alpha, \beta, \chi, \eta, \nu$ satisfy the corresponding equations
(\ref{lagmul},\ref{nueq},\ref{albetaeq}) must satisfy Euler's equations.
Let us calculate the material derivative of $\vec v$:
\beq
\frac{d\vec v}{dt} = \frac{d\vec \nabla \nu}{dt}  + \frac{d\alpha}{dt} \vec \nabla \chi +
 \alpha \frac{d\vec \nabla \chi}{dt}  +
\frac{d\beta}{dt} \vec \nabla \eta + \beta \frac{d\vec \nabla \eta}{dt}.
\label{dvform}
\enq
It can be easily shown that:
\ber
\frac{d\vec \nabla \nu}{dt} & = & \vec \nabla \frac{d \nu}{dt}- \vec \nabla v_k \frac{\partial \nu}{\partial x_k}
 = \vec \nabla (\frac{1}{2} \vec v^2 - w)- \vec \nabla v_k \frac{\partial \nu}{\partial x_k},
 \nonumber \\
 \frac{d\vec \nabla \eta}{dt} & = & \vec \nabla \frac{d \eta}{dt}- \vec \nabla v_k \frac{\partial \eta}{\partial x_k}
 = - \vec \nabla v_k \frac{\partial \eta}{\partial x_k},
 \nonumber \\
 \frac{d\vec \nabla \chi}{dt} & = & \vec \nabla \frac{d \chi}{dt}- \vec \nabla v_k \frac{\partial \chi}{\partial x_k}
 = - \vec \nabla v_k \frac{\partial \chi}{\partial x_k}.
 \label{dnabla}
\enr
In which $x_k$ is a Cartesian coordinate and a summation convention is assumed. Equations (\ref{lagmul},\ref{nueq})
where used in the above derivation. Inserting the result from equations (\ref{dnabla},\ref{albetaeq})
into \ern{dvform} yields:
\ber
\frac{d\vec v}{dt} &=& - \vec \nabla v_k (\frac{\partial \nu}{\partial x_k} + \alpha \frac{\partial \chi}{\partial x_k} +
\beta \frac{\partial \eta}{\partial x_k}) + \vec \nabla (\frac{1}{2} \vec v^2 - w)
 \nonumber \\
&+& \frac{1}{\rho} ((\vec \nabla \eta \cdot \vec J)\vec \nabla \chi - (\vec \nabla \chi \cdot \vec J)\vec \nabla \eta)
 \nonumber \\
&=& - \vec \nabla v_k v_k + \vec \nabla (\frac{1}{2} \vec v^2 - w)
 + \frac{1}{\rho} \vec J \times (\vec \nabla \chi \times  \vec \nabla \eta)
 \nonumber \\
&=& - \frac{\vec \nabla p}{\rho} + \frac{1}{\rho} \vec J \times \vec B.
\label{dvform2}
\enr
In which we have used both \ern{vform} and \ern{Bsakurai2} in the above derivation. This of course
proves that the barotropic Euler equations can be derived from the action given in \er{Lagactionsimp} and hence
all the equations of barotropic magnetohydrodynamics can be derived from the above action
without restricting the variations in any way except on the relevant boundaries and cuts.
The reader should take into account that the topology of the magnetohydrodynamic flow is conserved,
hence cuts must be introduced into the calculation as initial conditions.

\subsection{Simplified action}

The reader of this paper might argue here that the paper is misleading. The authors have declared
that they are going to present a simplified action for barotropic magnetohydrodynamics instead they
have added five more functions $\alpha,\beta,\chi,\eta,\nu$ to the standard set $\vec B,\vec v,\rho$.
In the following we will show that this is not so and the action given in \ern{Lagactionsimp} in
a form suitable for a pedagogic presentation can indeed be simplified. It is easy to show
that the Lagrangian density appearing in \ern{Lagactionsimp} can be written in the form:
\ber
{\cal L} & = & -\rho [\frac{\partial{\nu}}{\partial t} + \alpha \frac{\partial{\chi}}{\partial t}
+ \beta \frac{\partial{\eta}}{\partial t}+\varepsilon (\rho)] +
\frac{1}{2}\rho [(\vec v-\hat{\vec v})^2-(\hat{\vec v})^2]
\nonumber \\
& + &   \frac{1}{8 \pi} [(\vec B-\hat{\vec B})^2-(\hat{\vec B})^2]+
\frac{\partial{(\nu \rho)}}{\partial t} + \vec \nabla \cdot (\nu \rho \vec v ).
\label{Lagactionsimp4}
\enr
In which $\hat{\vec v}$ is a shorthand notation for $\vec \nabla \nu + \alpha \vec \nabla \chi +
 \beta \vec \nabla \eta$ (see \ern{vform}) and $\hat{\vec B}$ is a shorthand notation for
 $\vec \nabla \chi \times \vec \nabla \eta$ (see \ern{Bsakurai2}). Thus ${\cal L}$ has four contributions:
\ber
{\cal L} & = & \hat {\cal L} + {\cal L}_{\vec v}+ {\cal L}_{\vec B}+{\cal L}_{boundary},
\nonumber \\
\hat {\cal L} &\equiv & -\rho \left[\frac{\partial{\nu}}{\partial t} + \alpha \frac{\partial{\chi}}{\partial t}
+ \beta \frac{\partial{\eta}}{\partial t}+\varepsilon (\rho)+
\frac{1}{2} (\vec \nabla \nu + \alpha \vec \nabla \chi +  \beta \vec \nabla \eta)^2 \right]
\nonumber \\
&-&\frac{1}{8 \pi}(\vec \nabla \chi \times \vec \nabla \eta)^2
\nonumber \\
{\cal L}_{\vec v} &\equiv & \frac{1}{2}\rho (\vec v-\hat{\vec v})^2,
\nonumber \\
{\cal L}_{\vec B} &\equiv & \frac{1}{8 \pi} (\vec B-\hat{\vec B})^2,
\nonumber \\
{\cal L}_{boundary} &\equiv & \frac{\partial{(\nu \rho)}}{\partial t} + \vec \nabla \cdot (\nu \rho \vec v ).
\label{Lagactionsimp5}
\enr
The only term containing $\vec v$ is\footnote{${\cal L}_{boundary}$ also depends on
$\vec v$ but being a boundary term is space and time it does not contribute to the derived equations}
 ${\cal L}_{\vec v}$, it can easily be seen that
this term will lead, after we nullify the variational derivative with respect to $\vec v$,
to \ern{vform} but will otherwise
have no contribution to other variational derivatives. Similarly the only term containing $\vec B$
is ${\cal L}_{\vec B}$ and it can easily be seen that
this term will lead, after we nullify the variational derivative, to \ern{Bsakurai2} but will
have no contribution to other variational derivatives. Also notice that the term ${\cal L}_{boundary}$
contains only complete partial derivatives and thus can not contribute to the equations although
it can change the boundary conditions. Hence we see that \ers{lagmul}, \er{nueq} and \ers{albetaeq}
can be derived using the Lagrangian density $\hat {\cal L}[\alpha,\beta,\chi,\eta,\nu,\rho]$
in which $\hat{\vec v}$ replaces $\vec v$ and $\hat{\vec B}$ replaces $\vec B$ in the relevant equations.
Furthermore, after integrating the six \eqs
(\ref{lagmul},\ref{nueq},\ref{albetaeq}) we can insert the potentials $\alpha,\beta,\chi,\eta,\nu$
into \ers{vform} and (\ref{Bsakurai2}) to obtain the physical quantities $\vec v$ and $\vec B$.
Hence, the general barotropic magnetohydrodynamic problem is reduced from seven equations
(\ref{Beq},\ref{masscon},\ref{Euler}) and the additional constraint (\ref{Bcon})
to a problem of six first order (in the temporal derivative) unconstrained equations.
Moreover, the entire set of equations can be derived from the Lagrangian density $\hat {\cal L}$
which is what we were aiming to prove.

\subsection{The inverse problem}
\label{inverse}

In the previous subsection we have shown that given a set of functions $\alpha,\beta,\chi,\eta,\nu$
satisfying the set of equations described in the previous subsections,
one can insert those functions into \ern{vform} and \ern{Bsakurai2} to obtain the physical
velocity $\vec v$ and magnetic field $\vec B$. In this subsection we will address the inverse problem
that is, suppose we are given the quantities $\vec v,\vec B$ and $\rho$ how can one calculate the
potentials $\alpha,\beta,\chi,\eta,\nu$? The treatment in this section will follow closely
an analogue treatment for non-magnetic fluid dynamics given by Lynden-Bell \& Katz \cite{LynanKatz}.

\begin{figure}
\vspace{8cm}
\includegraphics{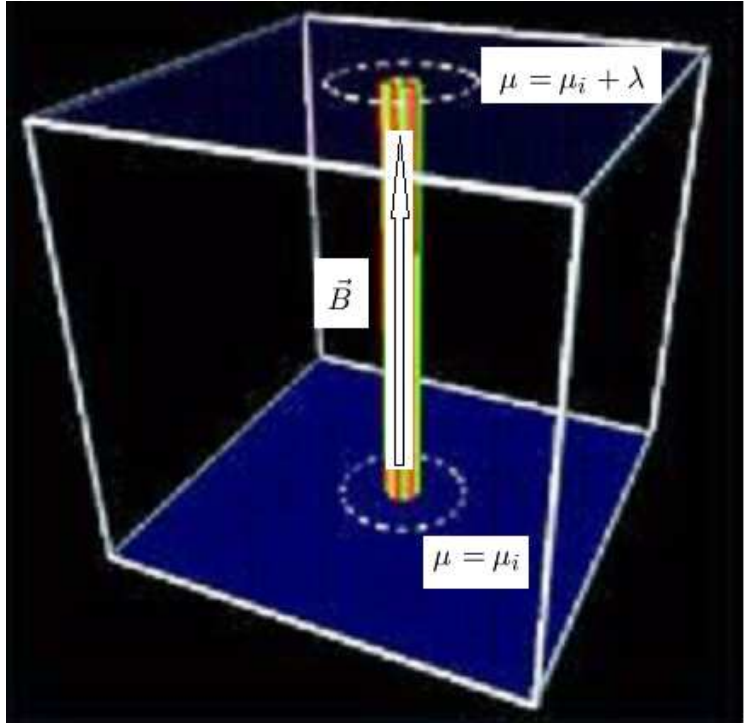}
\caption {A thin tube surrounding a magnetic field line}
\label{load}
\end{figure}

Consider a thin tube surrounding a magnetic field line as described in figure \ref{load},
the magnetic flux contained within the tube is:
\beq
\Delta \Phi =
\int \vec B \cdot d \vec S
\label{flux}
\enq
and the mass
contained with the tube is:
\beq
\Delta M = \int \rho d\vec l \cdot d \vec S,
\label{Mass}
\enq
in which $dl$ is a length element
along the tube. Since the magnetic field lines move with the flow
by virtue of \ern{Beq} both the quantities $\Delta \Phi$ and
$\Delta M$ are conserved and since the tube is thin we may define
the conserved magnetic load:
\beq
\lambda = \frac{\Delta M}{\Delta
\Phi} = \oint \frac{\rho}{B}dl,
\label{Load}
\enq
in which the
above integral is performed along the field line. Obviously the
parts of the line which go out of the flow to regions in which
$\rho=0$ have a null contribution to the integral.
Notice that $\lambda$ is a {\bf single valued} function that can be measured in principle.
Since $\lambda$
is conserved it satisfies the equation:
\beq
 \frac{d \lambda }{d t} = 0.
\label{Loadcon}
\enq
By construction surfaces of constant magnetic load move with the flow and contain
magnetic field lines. Hence the gradient to such surfaces must be orthogonal to
the field line:
\beq
\vec \nabla \lambda \cdot \vec B = 0.
\label{Loadortho}
\enq
Now consider an arbitrary comoving point on the magnetic field line and denote it by $i$,
and consider an additional comoving point on the magnetic field line and denote it by $r$.
The integral:
\beq
\mu(r)  = \int_i^r \frac{\rho}{B}dl + \mu(i),
\label{metage}
\enq
is also a conserved quantity which we may denote following Lynden-Bell \& Katz \cite{LynanKatz}
as the magnetic metage. $\mu(i)$ is an arbitrary number which can be chosen differently for each
magnetic line. By construction:
\beq
 \frac{d \mu }{d t} = 0.
\label{metagecon}
\enq
Also it is easy to see that by differentiating along the magnetic field line we obtain:
\beq
 \vec \nabla \mu \cdot \vec B = \rho.
\label{metageeq}
\enq
Notice that $\mu$ will be generally a {\bf non single valued} function, we will show later in this paper
that symmetry to translations in $\mu$ will generate through the Noether theorem the conservation of the
magnetic cross helicity.

\begin{figure}
\vspace{8cm}
\includegraphics{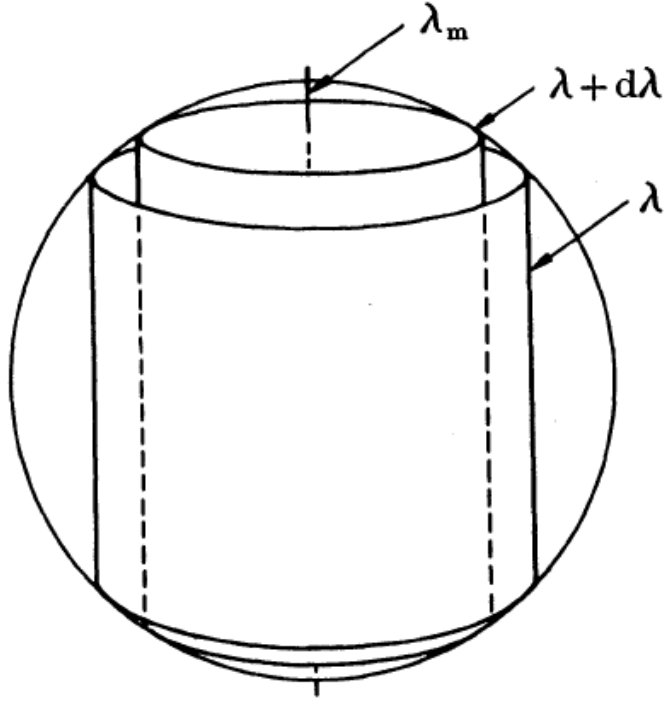}
\caption {Surfaces of constant load}
\label{loadsurface}
\end{figure}

At this point we have two comoving coordinates of flow, namely $\lambda,\mu$ obviously in a
three dimensional flow we also have a third coordinate. However, before defining the third coordinate
we will find it useful to work not directly with $\lambda$ but with a function of $\lambda$.
Now consider the magnetic flux within a surface of constant load $\Phi(\lambda)$ as described in figure \ref{loadsurface}
(the figure was given by Lynden-Bell \& Katz \cite{LynanKatz}). The magnetic flux is a conserved quantity
and depends only on the load $\lambda$ of the surrounding surface. Now we define the quantity:
\beq
 \chi = \frac{\Phi(\lambda)}{2\pi}.
\label{chidef}
\enq
Obviously $\chi$ satisfies the equations:
\beq
\frac{d \chi}{d t} = 0, \qquad \vec B \cdot \vec \nabla \chi = 0,
\label{chieq}
\enq
we will immediately show that this function is identical to Sakurai's function
defined in \ern{Bsakurai}. Let us now define an additional comoving coordinate $\eta^{*}$
since $\vec \nabla \mu$ is not orthogonal to the $\vec B$ lines we can choose $\vec \nabla \eta^{*}$ to be
orthogonal to the $\vec B$ lines and not be in the direction of the $\vec \nabla \chi$ lines,
that is we choose $\eta^{*}$
not to depend only on $\chi$. Since both $\vec \nabla \eta^{*}$ and $\vec \nabla \chi$ are orthogonal to $\vec B$,
$\vec B$ must take the form:
\beq
\vec B = A \vec \nabla \chi \times \vec \nabla \eta^{*}.
\enq
However, using \ern{Bcon} we have:
\beq
\vec \nabla \cdot \vec B = \vec \nabla A \cdot (\vec \nabla \chi \times \vec \nabla \eta^{*})=0.
\enq
Which implies that $A$ is a function of $\chi,\eta^{*}$. Now we can define a new comoving function
$\eta$ such that:
\beq
\eta = \int_0^{\eta^{*}}A(\chi,\eta^{'*})d\eta^{'*}, \qquad \frac{d \eta}{d t} = 0.
\enq
In terms of this function we recover the Sakurai presentation defined in \ern{Bsakurai}:
\beq
\vec B = \vec \nabla \chi \times \vec \nabla \eta.
\label{Bsakurai3}
\enq
Hence we have shown how $\chi,\eta$ can be constructed for a known $\vec B,\rho$.
Notice however, that $\eta$ is defined in a non unique way since one can redefine
$\eta$ for example by performing the following transformation: $\eta \rightarrow \eta + f(\chi)$
in which $f(\chi)$ is an arbitrary function.
The comoving coordinates $\chi,\eta$ serve as labels of the magnetic field lines.
Moreover the magnetic flux can be calculated as:
\beq
\Phi = \int \vec B \cdot d \vec S = \int d \chi d \eta.
\label{phichieta}
\enq
In the case that the surface integral is performed inside a load contour we
obtain:
\beq
\Phi (\lambda) = \int_{\lambda} d \chi d \eta= \chi \int_{\lambda} d \eta =\left\{
\begin{array}{c}
 \chi [\eta] \\
  \chi (\eta_{max}-\eta_{min}) \\
\end{array}
\right.
\enq
Comparing the above \eq \ with \ern{chidef} we derive that $\eta$ can be either
{\bf single valued} or {\bf not single valued} and that
its discontinuity across its cut in the non single valued case is $[\eta] =2 \pi$.

We will now show how the potentials $\alpha,\beta,\nu$ can be derived.
Let us calculate the vorticity $\vec \omega$ of the flow. By taking the curl of
\ern{vform} we obtain:
\beq
\vec \omega \equiv \vec \nabla \times \vec v =
 \vec \nabla \alpha  \times \vec \nabla \chi + \vec \nabla \beta \times \vec \nabla \eta.
\label{vort}
\enq
The following identities are derived:
\beq
\vec \omega \cdot \vec \nabla \chi = (\vec \nabla \beta \times \vec \nabla \eta) \cdot \vec \nabla \chi=
- \vec \nabla \beta \cdot \vec B,
\label{omchi}
\enq
\beq
\vec \omega \cdot \vec \nabla \eta = (\vec \nabla \alpha \times \vec \nabla \chi) \cdot \vec \nabla \eta=
\vec \nabla \alpha \cdot \vec B.
\label{ometa}
\enq
Now let us perform integrations along $\vec B$ lines starting from an arbitrary point denoted as $i$
to another arbitrary point denoted as $r$.
\beq
\beta (r) = - \int_i^r \frac{\vec \omega \cdot \vec \nabla \chi}{B} dl + \beta(i),
\label{betadef}
\enq
\beq
\alpha (r) =  \int_i^r \frac{\vec \omega \cdot \vec \nabla \eta}{B} dl + \alpha(i).
\label{alphadef}
\enq
The numbers $\alpha(i),\beta(i)$ can be chosen in an arbitrary way for each magnetic field line.
Hence we have derived (in a non-unique way) the values of the $\alpha,\beta$ functions.
Finally we can use \ern{vform} to derive the function $\nu$ for any point $s$ within the flow:
\beq
\nu(s) = \int_i^s (\vec v - \alpha \vec \nabla \chi - \beta \vec \nabla \eta)\cdot d \vec r + \nu(i),
\enq
in which $i$ is any arbitrary point within the flow, the result will not depend on the trajectory taken
in the case that $\nu$ is single valued. If $\nu$ is not single valued on should introduce a cut
which the integration trajectory should not cross.

\subsection{Stationary barotropic magnetohydrodynamics}
\label{Statmag}

Stationary flows are a unique phenomena of Eulerian fluid dynamics which has
no counter part in Lagrangian fluid dynamics. The stationary flow is defined
by the fact that the physical fields $\vec v,\vec B,\rho$ do not depend on the
temporal coordinate. This, however, does not imply that the corresponding potentials
$\alpha,\beta,\chi,\eta,\nu$ are all functions of spatial coordinates alone.
Moreover, it can be shown that choosing the potentials in such a way will lead
to erroneous results in the sense that the stationary equations of motion can
not be derived from the Lagrangian density $\hat {\cal L}$ given in \ern{Lagactionsimp5}.
However, this problem can be amended easily
as follows. Let us choose $\alpha,\beta,\chi,\nu$ to depend on the spatial coordinates alone.
Let us choose $\eta$ such that:
\beq
\eta = \bar \eta - t,
\label{etastas}
\enq
in which $\bar \eta$ is a function of the spatial coordinates. The Lagrangian
density $\hat {\cal L}$ given in \ern{Lagactionsimp5} will take the form:
\beq
\hat {\cal L} = \rho (\beta -\varepsilon (\rho))-
\frac{1}{2}\rho (\vec \nabla \nu + \alpha \vec \nabla \chi + \beta \vec \nabla \bar \eta)^2
-\frac{1}{8 \pi}(\vec \nabla \chi \times \vec \nabla \bar \eta)^2.
\label{stathatL}
\enq
The above functional can be compared
with Vladimirov and Moffatt \cite{Moffatt} equation 6.12 for incompressible flows in which their
$I$ is analogue to our $\beta$. Notice however, that while $\beta$ is not a conserved quantity $I$ is.

Varying the Lagrangian $\hat {L} = \int \hat {\cal L} d^3x$ with respect to $\nu,\alpha,\beta,\chi,\eta,\rho$
leads to the following equations:
\ber
& &  \vec \nabla \cdot (\rho \hat {\vec v} ) = 0,
\nonumber \\
& & \rho \hat {\vec v} \cdot \vec \nabla \chi = 0,
\nonumber \\
& & \rho (\hat {\vec v} \cdot \vec \nabla \bar \eta - 1) = 0,
\nonumber \\
& & \hat {\vec v} \cdot \vec \nabla \alpha = \frac{\vec \nabla \bar \eta \cdot \hat{\vec J}}{\rho}, \qquad
\nonumber \\
& & \hat {\vec v} \cdot \vec \nabla \beta = -\frac{\vec \nabla \chi \cdot \hat{\vec J}}{\rho},
\nonumber \\
& & \beta= \frac{1}{2} \hat {\vec v}^2 + w.
\label{statlagmul}
\enr
Calculations similar to the ones done in previous subsections will show that those equations
lead to the stationary barotropic magnetohydrodynamic equations:
\beq
 \vec \nabla \times (\hat {\vec v} \times \hat{\vec B}) = 0,
\label{Beqstat}
\enq
\beq
\rho (\hat {\vec v} \cdot \vec \nabla) \hat {\vec v} = -\vec \nabla p (\rho) +
\frac{(\vec \nabla \times \hat{\vec B}) \times \hat{\vec B}}{4 \pi}.
\label{Eulerstat}
\enq

\section{The Simplified Hamiltonian Formalism}

Let us derive the conjugate momenta of the variables appearing in the Lagrangian density
$\hat {\cal L}$ defined in \ern{Lagactionsimp5}. A simple calculation will yield:
\beq
\pi_\nu \equiv \frac{\partial \hat {\cal L}}{\partial\left(\frac{\partial{\nu}}{\partial t}\right)} = -\rho, \quad
\pi_\chi \equiv \frac{\partial \hat {\cal L}}{\partial\left(\frac{\partial{\chi}}{\partial t}\right)} = -\rho \alpha, \quad
\pi_\eta \equiv \frac{\partial \hat {\cal L}}{\partial\left(\frac{\partial{\eta}}{\partial t}\right)} = -\rho \beta.
\label{conjugamomenta}
\enq
The rest of the canonical momenta $\pi_\rho, \pi_\alpha, \pi_\beta$ are null. It thus seems that the six functions
appearing in the Lagrangian density $\hat {\cal L}$ can be divided to "approximate" conjugate pairs:
$(\nu,\rho), (\chi, \alpha), (\eta, \beta)$. The Hamiltonian density $\hat {\cal H}$ can be now calculated as
follows:
\beq
\hat {\cal H} = \pi_\nu \frac{\partial{\nu}}{\partial t} + \pi_\chi \frac{\partial{\chi}}{\partial t}
+ \pi_\eta \frac{\partial{\eta}}{\partial t} - \hat {\cal L} =
\rho \left[\varepsilon (\rho)+ \frac{1}{2} \hat{\vec v}^2 \right]
+\frac{1}{8 \pi} \hat{\vec B}^2,
\label{hamiltonian}
\enq
in which $\hat{\vec v}$ is defined in \ern{vform} and $\hat{\vec B}$ is defined in \ern{Bsakurai2}.
This Hamiltonian was previously introduced by Frenkel, Levich \& Stilman \cite{FLS} using somewhat
different variables\footnote{The following notations are used in \cite{FLS}: $\lambda=\chi, \Lambda=\eta, \mu= \pi_\chi
, M=\pi_\eta, \phi=\nu$}. The equations derived from the above Hamiltonian density are
similar to \ers{lagmul}, \er{nueq} and \ers{albetaeq} and will not be re-derived here.
While Frenkel, Levich \& Stilman \cite{FLS} have postulated the Hamiltonian density appearing
in \ern{hamiltonian}, this Hamiltonian is here derived from a Lagrangian.

\section{Simplified variational principle of stationary barotropic magnetohydrodynamics}

In the previous section we have shown that barotropic magnetohydrodynamics
can be described in terms of six first order differential equations or
 of an action principle from which those equations can be derived.
This formalism was shown to apply to both stationary and non-stationary magnetohydrodynamics.
Although for non-stationary magnetohydrodynamics we do not know at present how the
number of functions can be further reduced, for stationary barotropic magnetohydrodynamics the
situation is quite different. In fact we will show that for stationary barotropic magnetohydrodynamics
three functions will suffice.

Consider \ern{chieq}, for a stationary flow it takes the form:
\beq
\vec v \cdot \vec \nabla \chi = 0.
\label{chieqstat}
\enq
Hence $\vec v$ can take the form:
\beq
\vec v = \frac{\vec \nabla \chi \times \vec K}{\rho}.
\label{orthov}
\enq
However, since the velocity field must satisfy the stationary mass conservation \ern{masscon}:
\beq
\vec \nabla \cdot (\rho \vec v ) = 0.
\label{massconstat}
\enq
We see that $\vec K$ must have the form $\vec K = \vec \nabla N$, where $N$ is an arbitrary function. Thus,
$\vec v$ takes the form:
\beq
\vec v = \frac{\vec \nabla \chi \times \vec \nabla N}{\rho}.
\label{orthov2}
\enq
Let us now calculate $\vec v \times \vec B$ in which $\vec B$ is given by Sakurai's presentation
\ern{Bsakurai2}:
\ber
\vec v \times \vec B &=& (\frac{\vec \nabla \chi \times \vec \nabla N}{\rho}) \times
(\vec \nabla \chi \times \vec \nabla \eta)
\nonumber \\
&=& \frac{1}{\rho} \vec \nabla \chi (\vec \nabla \chi \times \vec \nabla N) \cdot \vec \nabla \eta.
\label{orthov3}
\enr
Since the flow is stationary $N$ can be at most a function of the three comoving coordinates
$\chi,\mu,\bar \eta$ defined in subsections \ref{inverse} and \ref{Statmag}, hence:
\beq
\vec \nabla N  = \frac{\partial N}{\partial \chi} \vec \nabla \chi +
\frac{\partial N}{\partial \mu} \vec \nabla \mu + \frac{\partial N}{\partial \bar \eta} \vec \nabla \bar \eta.
\label{Ndiv}
\enq
Inserting \ern{Ndiv} into \er{orthov3} will yield:
\beq
\vec v \times \vec B =
\frac{1}{\rho} \vec \nabla \chi \frac{\partial N}{\partial \mu}
(\vec \nabla \chi \times \vec \nabla \mu) \cdot \vec \nabla \bar \eta.
\label{orthov4}
\enq
Rearranging terms and using Sakurai's presentation \ern{Bsakurai2} we can
simplify the above equation and obtain:
\beq
\vec v \times \vec B = -\frac{1}{\rho} \vec \nabla \chi \frac{\partial N}{\partial \mu}
(\vec \nabla \mu \cdot \vec B).
\label{orthov5}
\enq
However, using \ern{metagecon} this will simplify to the form:
\beq
\vec v \times \vec B = - \vec \nabla \chi \frac{\partial N}{\partial \mu}.
\label{orthov51}
\enq
Now let us consider \ern{Beq}; for stationary flows this will take the form:
\beq
\vec \nabla \times (\vec v \times \vec B) = 0.
\label{Beqstat2}
\enq
Inserting \ern{orthov5} into \ern{Beqstat} will lead to the equation:
\beq
\vec \nabla  (\frac{\partial N}{\partial \mu}) \times \vec \nabla \chi = 0.
\label{Beqstatimp}
\enq
However, since $N$ is at most a function of $\chi,\mu,\bar \eta$ it follows that
$\frac{\partial N}{\partial \mu}$ is some function of $\chi$:
\beq
\frac{\partial N}{\partial \mu} = -F(\chi).
\enq
This can be easily integrated to yield:
\beq
N = - \mu F(\chi) + G(\chi,\bar \eta).
\enq
Inserting this back into \ern{orthov2} will yield:
\beq
\vec v = \frac{\vec \nabla \chi \times (-F(\chi) \vec \nabla \mu+
\frac{\partial G}{\partial \bar \eta} \vec \nabla \bar \eta) }{\rho}.
\label{orthov7}
\enq
Let us now replace the set of variables $\chi,\bar \eta$ with a new set $\chi',\bar \eta'$
such that:
\beq
\chi' = \int F(\chi) d\chi, \qquad \bar \eta' = \frac{\bar \eta}{F(\chi)}.
\enq
This will not have any effect on the Sakurai representation given in \ern{Bsakurai2} since:
\beq
\vec B = \vec \nabla \chi \times \vec \nabla \eta = \vec \nabla \chi \times \vec \nabla \bar \eta =
\vec \nabla \chi' \times \vec \nabla \bar \eta'.
\enq
However, the velocity will have a simpler representation and will take the form:
\beq
\vec v = \frac{\vec \nabla \chi' \times \vec \nabla(- \mu+ G'(\chi',\bar \eta'))}{\rho},
\label{orthov8}
\enq
in which $G'=\frac{G}{F}$. At this point one should remember that $\mu$ was defined in \ern{metage}
up to an arbitrary constant which can vary between magnetic field lines. Since the lines
are labelled by their $\chi',\bar \eta'$ values it follows that we can add an arbitrary function of
$\chi',\bar \eta'$ to $\mu$ without effecting its properties. Hence we can define a new $\mu'$ such that:
\beq
\mu' =\mu-  G'(\chi',\bar \eta').
\label{mupdef}
\enq
Notice that $\mu'$ can be multi-valued; this will be discussed in somewhat more detail
in subsection \ref{torussec}.
Inserting \er{mupdef} into \er{orthov8} will lead to a simplified equation for $\vec v$:
\beq
\vec v = \frac{\vec \nabla \mu' \times \vec \nabla \chi'}{\rho}.
\label{orthov9}
\enq
In the following the primes on $\chi,\mu,\bar \eta$ will be ignored.
The above equation is analogues to Vladimirov and Moffatt's \cite{Moffatt} equation 7.11
for incompressible flows, in which our $\mu$ and $\chi$ play the part of their $A$ and
$\Psi$. It is obvious that
$\vec v$ satisfies the following set of equations:
\beq
\vec v \cdot \vec \nabla \mu = 0, \qquad  \vec v \cdot \vec \nabla \chi = 0, \qquad
\vec v \cdot \vec \nabla \bar \eta = 1,
\label{comov}
\enq
to derive the right hand equation we have used both \ern{metagecon} and \ern{Bsakurai2}.
Hence $\mu,\chi$ are both comoving and stationary. As for $\bar \eta$ it satisfies the same
equation as $\bar \eta$ defined in \ern{etastas} as can be seen from \ern{statlagmul}.
It can be easily seen that if:
\beq
basis = (\vec \nabla \chi, \vec \nabla \bar \eta,\vec \nabla \mu),
\enq
is a local vector basis at any point in space than their exists a dual basis:
\beq
dual \ basis = \frac{1}{\rho}(\vec \nabla \bar \eta \times \vec \nabla \mu, \vec \nabla \mu \times \vec \nabla \chi
, \vec \nabla \chi \times \vec \nabla \bar \eta ) =
(\frac{\vec \nabla \bar \eta \times \vec \nabla \mu}{\rho}, \vec v, \frac{\vec B}{\rho} ).
\enq
Such that:
\beq
basis_i \cdot dual \ basis_j = \delta_{ij}, \qquad i,j\in[1,2,3],
\enq
in which $\delta_{ij}$ is Kronecker's delta.
Hence while the surfaces $\chi,\mu,\bar \eta$ generate a local vector basis for space, the
physical fields of interest $\vec v,\vec B$ are part of the dual basis.
By vector multiplying $\vec v$ and $\vec B$ and using equations (\ref{orthov9},\ref{Bsakurai2})
we obtain:
\beq
\vec v \times \vec B = \vec \nabla \chi,
\label{vBsurfaces}
\enq
this means that both $\vec v$ and $\vec B$ lie on $\chi$ surfaces and provide a vector
basis for this two dimensional surface. The above equation can be compared
with Vladimirov and Moffatt \cite{Moffatt} equation 5.6 for incompressible flows in which their
$J$ is analogue to our $\chi$.

\subsection{The action principle}

In the previous subsection we have shown that if the velocity field $\vec v$
is given by \ern{orthov9} and the magnetic field $\vec B$ is given by the Sakurai
representation \ern{Bsakurai2} than \eqs (\ref{Beq},\ref{Bcon},\ref{masscon})
are satisfied automatically for stationary flows. To complete the set of equations
we will show how the Euler \ers{Euler} can be derived from the action given in \ern{Lagaction}
in which both $\vec v$ and $\vec B$ are given by \ern{orthov9} and \ern{Bsakurai2} respectively
and the density $\rho$ is given by \ern{metagecon}:
\beq
\rho = \vec \nabla \mu \cdot \vec B = \vec \nabla \mu \cdot (\vec \nabla \chi \times \vec \nabla \eta)
=\frac{\partial(\chi,\eta,\mu)}{\partial(x,y,z)}.
\label{metagecon2}
\enq
In this case the Lagrangian density of \ern{Lagaction} will take the form:
\beq
{\cal L} = \rho (\frac{1}{2} (\frac{\vec \nabla \mu \times \vec \nabla \chi}{\rho})^2 - \varepsilon (\rho)) -
\frac{(\vec \nabla \chi \times \vec \nabla \eta)^2}{8 \pi}
\label{lagstatsimp}
\enq
and can be seen explicitly to depend on only three functions.
Let us make arbitrary small variations $\delta \alpha_i =(\delta \chi,\delta \eta,\delta \mu)$ of the functions
$\alpha_i=(\chi,\eta,\mu)$. Let us define the vector:
\beq
\vec \xi \equiv -\frac{\partial \vec r}{\partial \alpha_i} \delta \alpha_i.
\label{xialdef}
\enq
This will lead to the equation:
\beq
\delta \alpha_i = -\vec \nabla \alpha_i \cdot \vec \xi.
\label{delal}
\enq
And by virtue of \ern{delF} we have:
\beq
\Delta \alpha_i = \delta \alpha_i + (\vec \xi \cdot \vec \nabla) \alpha_i = 0,
\label{Delal}
\enq
as one should expect since $\alpha_i$ are comoving with the flow. Making a variation
of $\rho$ given in \ern{metagecon2} with respect to $\alpha_i$ will yield \ern{delrho}.
Furthermore, taking the variation of $\vec B$ given by Sakurai's representation
(\ref{Bsakurai2}) with respect to $\alpha_i$ will yield
\ern{delB}. It remains to
calculate $\delta \vec v$ by varying \ern{orthov9} this will yield:
\beq
\delta \vec v = -\frac{\delta \rho}{\rho} \vec v + \frac{1}{\rho} \vec \nabla \times (\rho \vec \xi \times \vec v).
\label{delv2}
\enq
Inserting \eqs (\ref{delrho},\ref{delB},\ref{delv2}) into \ern{delLagaction}
will yield:
\ber
\delta {\cal L} &=& \vec v \cdot \vec \nabla \times (\rho \vec \xi \times \vec v)
- \frac{\vec B \cdot \vec \nabla \times (\vec \xi \times \vec B)}{4 \pi}
- \delta \rho (\frac{1}{2} \vec v^2 + w)
\nonumber \\
&=& \vec v \cdot \vec \nabla \times (\rho \vec \xi \times \vec v)
- \frac{\vec B \cdot \vec \nabla \times (\vec \xi \times \vec B)}{4 \pi}
+ \vec \nabla \cdot (\rho \vec \xi) (\frac{1}{2} \vec v^2 + w ).
\label{delcalL}
\enr
Using the well known vector identity:
\beq
\vec A \cdot \vec \nabla \times (\vec C \times \vec A)=
\vec \nabla \cdot ((\vec C \times \vec A) \times \vec A)+
(\vec C \times \vec A) \cdot \vec \nabla \times \vec A
\label{veciden1}
\enq
and the theorem of Gauss we can write now \ern{delLagaction} in the form:
\ber
\delta A & = &
\int dt \{ \oint d \vec S \cdot [\rho (\vec \xi \times \vec v)\times \vec v
-\frac{(\vec \xi \times \vec B)\times \vec B}{4 \pi}+(\frac{1}{2} \vec v^2 + w )\rho \vec \xi]
\nonumber \\
& + & \int d^3 x  \vec \xi \cdot [\rho \vec v \times \vec \omega+
\vec J \times \vec B-\rho \vec \nabla (\frac{1}{2} \vec v^2 + w )]\}.
\label{delLagaction2simpl}
\enr
The time integration is of course redundant in the above expression. Also notice that we have used
the current definition \ern{J} and the vorticity definition \ern{vort}. Suppose now
that $\delta A = 0$ for a $\vec \xi$ such that the boundary term (including both the
boundary of the domain and relevant cuts) in the above equation
is null but that $\vec \xi$ is otherwise arbitrary, then it entails the equation:
\beq
\rho \vec v \times \vec \omega+
\vec J \times \vec B-\rho \vec \nabla (\frac{1}{2} \vec v^2 + w ) = 0.
\enq
Using the well known vector identity:
\beq
\frac{1}{2} \vec \nabla (\vec v^2) = (\vec v \cdot \vec \nabla) \vec v +
 \vec v \times (\vec \nabla \times \vec v)
\label{veciden2}
\enq
and rearranging terms we recover the stationary Euler equation:
\beq
\rho (\vec v \cdot \vec \nabla)\vec v = -\vec \nabla p +  \vec J \times \vec B.
\label{Eulerstat2}
\enq

\subsection {The case of an axi-symmetric magnetic field}

Consider an axi-symmetric magnetic field such that the magnetic
field is dependent only on the coordinate $R$ which is the distance
from the axis of symmetry and the coordinate $z$ which is the distance
along the axis of symmetry from an arbitrary origin on the axis.
Thus:
\beq
\vec B = \vec B (R,z).
\enq
Any axi-symmetric magnetic field
satisfying \er{Bcon} can be represented in the form:
\beq
\vec B = \vec \nabla P \times \vec \nabla (\frac{\phi}{2\pi}) + 2 \pi R B_{\phi} \vec \nabla (\frac{\phi}{2\pi}).
\label{BP}
\enq
In which $\phi$ is the azimuthal angle defined in the conventional way and $B_{\phi}$
is the component of $\vec B$ in the $\phi$ direction. The function $P=P (R,z)$ is
the flux through a circle of radius $R$ at height $z$:
\beq
P(R,z) = \int_{(R,z)} \vec B \cdot d \vec S  = 2 \pi \int_0^R B_z (R',z)  R' d R'.
\enq
For finite field configurations $P$ will have a maximum $P_m = P (R_m,z_m)$
at some $R_m,z_m$. This circle $R=R_m$ will form a line toroid with the other
constant $P$ surfaces nearby forming a nested set. There can be several such local
maxima with local nested toroids in a general configuration but the simpler case has
just one.

Let us study the relations between the functions $P,B_{\phi}$ and the functions
$\chi,\eta$ given in \ern{Bsakurai}. Assuming that the density $\rho$ is axi-symmetric
one can see the magnetic load defined in \ern{Load} is also axi-symmetric and that
the surfaces of constant load are surfaces of revolution around the axis of symmetry.
From \ern{chidef} we deduce that $\chi = \chi(R,z)$. Expressing \ern{Bsakurai} in terms
of the coordinates $R,\phi,z$ results in:
\beq
\vec B=(-\frac{1}{R} \partial_z \chi \partial_{\phi} \eta) \hat R
+ (\partial_z \chi \partial_R \eta - \partial_R \chi \partial_z \eta ) \hat \phi
+ (\frac{1}{R} \partial_R \chi \partial_{\phi} \eta) \hat z.
\label{BSaxsym}
\enq
In which $\partial_y$ is a short hand notation for $\frac{\partial}{\partial y}$
and $\hat y$ is a unit vector perpendicular to the constant $y$ surface.
Comparing \er{BSaxsym} with \er{BP} we arrive with the set of equations:
\beq
\partial_z P =\partial_z \chi \partial_{\phi} (2 \pi \eta), \qquad
\partial_R P =\partial_R \chi \partial_{\phi} (2 \pi \eta).
\label{Pchirel}
\enq
From which we derive the equation:
\beq
\partial_z P \partial_R \chi - \partial_R P \partial_z \chi = 0 \Rightarrow
\vec \nabla P \times \vec \nabla \chi = 0.
\enq
Hence:
\beq
P=F(\chi),
\enq
where $F$ is an arbitrary function. We deduce that $P$ is just another type
of labelling of the load surfaces.
Thus \ern{Pchirel} will lead to:
\beq
\partial_{\phi} (2 \pi \eta) = \frac{dP}{d\chi} \Rightarrow
\eta = \frac{\phi}{2 \pi}\frac{dP}{d\chi}+\tilde{\eta}(R,z).
\enq
This should be compared with the result of Young et al. \cite[equation 5.1]{Yang}.
Substituting the above result in \ern{BSaxsym} will lead to the equation:
\beq
B_{\phi} \hat \phi = (\partial_z \chi \partial_R \tilde{\eta} - \partial_R \chi \partial_z \tilde{\eta} ) \hat \phi
= \vec \nabla \chi \times \vec \nabla \tilde{\eta}.
\enq
This can also be written as:
\beq
B_{\phi} = \hat \phi \cdot (\vec \nabla \chi \times \vec \nabla \tilde{\eta})
= (\hat \phi \times \vec \nabla \chi) \cdot \vec \nabla \tilde{\eta}.
\enq
Hence $B_{\phi}$ is proportional to the gradient of $\tilde{\eta}$ along the $\hat \phi \times \vec \nabla \chi$
direction. Since $\hat \phi \times \vec \nabla \chi$ is known we can integrate along this vector
to obtain a non-unique solution for $\tilde{\eta}$:
\beq
\tilde{\eta} = \int \frac{B_{\phi}}{|\hat \phi \times \vec \nabla \chi|}dl,
\enq
in which $dl$ is a line element along the $\hat \phi \times \vec \nabla \chi$ line.

\subsection {The case of a magnetic field on a toroid}
\label{torussec}

Our previous definitions of the surfaces of constant load given in
\ern{Load} is ambiguous when the field lines are "surface filling"
eg on a toroid and give no result when the field lines are "volume
filling". At equilibrium $\vec B$ and $\vec v$ lie in surfaces
(there is an exception when $\vec B$ and $\vec v$ are parallel
and fill volumes). Our former considerations apply unchanged if
these surfaces have the topology of cylinders but they need generalization
when the surfaces have the topology of toroids nested on a line (a similar discussion
in which non-magnetic fluids are considered can be found in \cite{DLB}). We consider
a surface $\Sigma$ spanning that line toroid. Each toroid $T$ will meet $\Sigma$ in a loop.
Consider the magnetic flux $\Phi(T)$ through that part of $\Sigma$ within the loop
and the mass enclosed by the toroid $m(T)$. The the mass outside the toroid is
$\tilde{m}(T) = M - m(T)$. Now express $\tilde{m}$ as a function $\tilde{m}(\Phi)$
of the magnetic flux $\Phi$, then a definition of magnetic load analogous to that for
"cylinders" parallel to the axis is:
\beq
\lambda = \frac{d \tilde{m}}{d \Phi}.
\enq
However, there are now two loads corresponding to the two fluxes associated with a given toroid.
The other load is obtained by taking a cut across the "short" circle section of the torus say
of constant $\phi$. The magnetic flux $\Phi^{*}$ through such a cross section may be expressed
as a function of the total mass $m(T)$ within the toroid and
\beq
\lambda^{*} = \frac{d m}{d \Phi^{*}},
\enq
is a second different load. Of course it is also permissible to reexpress the flux $\Phi^{*}$ as
a function of the flux $\Phi$ then we find:
\beq
\lambda^{*} = \frac{d m}{d \Phi^{*}} = \frac{\frac{d m}{d \Phi}}{\frac{d \Phi^{*}}{d \Phi}}
= - \frac{\lambda}{\frac{d \Phi^{*}}{d \Phi}}.
\enq
The surfaces of constant $\lambda$ are of course the toroids $T$ which also have $\lambda^{*}$ constant.

A similar problem may arise with the definition of the magnetic metage defined in \ern{metage}.
We may wish to define this quantity using the magnetic field $\vec B$ and velocity field
$\vec v$. Since those vectors provide a vector basis on the load surface, they can
be combined in such a way say: $\vec B +\gamma (\chi) \vec v$ to create a vector which is
directed along the large loop of the toroid. (A different $\gamma$ will leave only twists
around the short way.)  This combination represents an unwinding of the field lines
so that they no longer twist around the short (long) way.
Those loops can be thought as composing the surface $\Sigma$.
Another surface $\Sigma'$ also composed of such untwisted loops can
be so chosen that the mass between $\Sigma$ and $\Sigma'$  and between
loads $\lambda$ and $\lambda+d\lambda$ is some fixed fraction of $\frac{dm}{d\lambda}d\lambda$.
Such $\Sigma'$ form suitable constant metage surfaces $\mu$ corresponding to partial loads
$\lambda$. Notice that $2 \pi \mu$ then describes the angle from $\Sigma$ turned around the
toroid by the short way to reach any chosen point. Similar use of the other load $\lambda^{*}$
allows us to define another generalized angle $\mu^{*}$ measured around the long way.
A somewhat less physical approach is given below.

Let us consider a toroid of constant magnetic load.  Dungey \cite[p. 31]{Dungey} has considered the case in
which magnetic field lines lie on a torus.
He has shown that one of the functions (ie $\eta$) involved in the representation (\ref{Bsakurai})
should be non-single valued and therefore a cut should be introduced.

\begin{figure}
\vspace{8cm}
\includegraphics{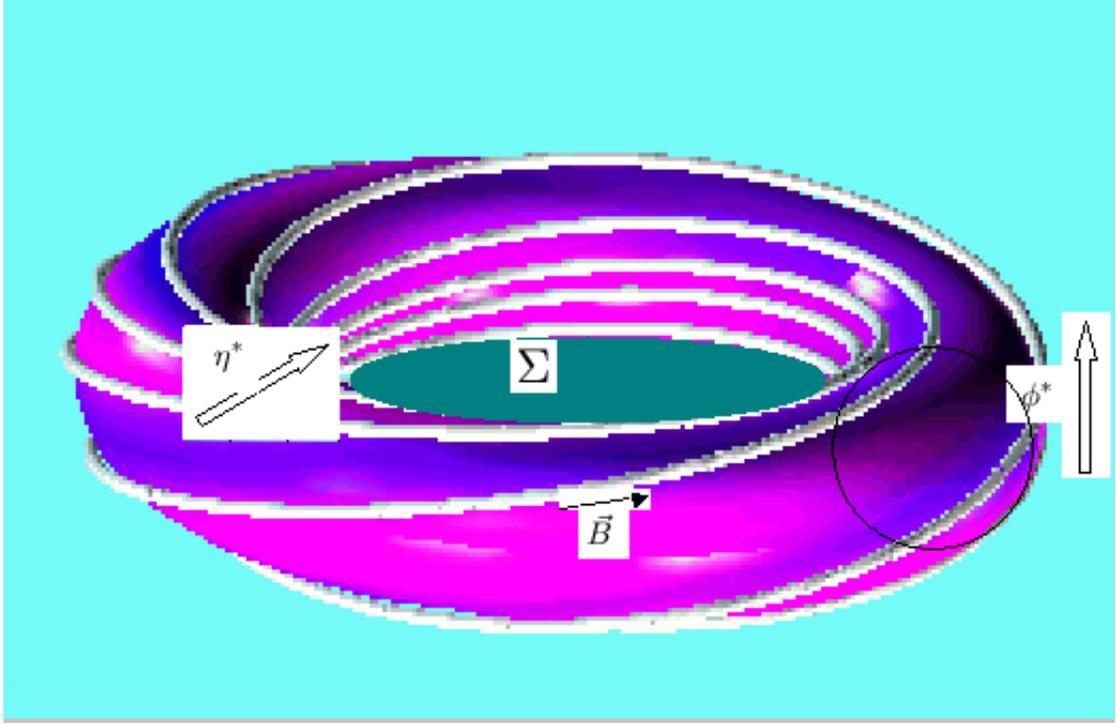}
\caption {A torus of magnetic field lines}
\label{torus}
\end{figure}

In order to obtain a simple
looking cuts we will replace the previous set of functions $\mu,\eta$ with a new set $\phi^{*},\eta^{*}$,
which will be defined as follows:
\ber
\phi^{*} &\equiv&  \frac{\mu + G(\chi) \eta}{\Omega(\chi)},
\nonumber \\
\eta^{*} &\equiv& \frac{\eta - \phi^{*}}{f(\chi)},
\label{phetstar}
\enr
$G,\Omega,f$ are arbitrary functions of $\chi$. Therefore $\mu$ and $\eta$ can be given
in terms of $\phi^{*},\eta^{*}$ as:
\ber
\eta &=&  \phi^{*} + f(\chi) \eta^{*},
\nonumber \\
\mu &=& -G(\chi) \eta + \Omega(\chi)\phi^{*},
\label{etmuintermstar}
\enr
$\eta^{*}$ can be considered as an angle varying over the small circle of the torus,
while $\phi^{*}$ can be considered as an angle varying over the large circle of the torus
as in figure (\ref{torus}). On the torus of constant magnetic load
the $\phi^{*},\eta^{*}$ functions have a simple "cut" structure.
The above equation can also serve as a "definition" of $\mu$.
Inserting \ers{etmuintermstar} into \ern{Bsakurai} and \ern{orthov9} will result
in the following set of equations:
\ber
\vec B & = & \vec \nabla \chi \times \vec \nabla \phi^{*} + f(\chi) \vec \nabla \chi \times \vec \nabla \eta^{*},
\nonumber \\
\vec v & = & G(\chi) \frac{\vec B}{\rho} +
 \Omega(\chi) \frac{\vec \nabla \phi^{*} \times \vec \nabla \chi}{\rho}.
\enr
Hence $\vec B$ is partitioned into two vectors circulating along the small and large
circles of the torus. While $\vec v$ has two vector components one along the magnetic field $\vec B$
and another along the small circle.

\section {Topological constants of motion}
\label{Top}

Magnetohydrodynamics is known to have the following two topological constants of motion;
one is the magnetic helicity:
\beq
{\cal H}_M \equiv \int \vec B \cdot \vec A d^3 x,
\label{maghel}
\enq
which is known to measure the degree of knottiness of lines of the magnetic field $\vec B$ \cite{Moff2}.
In the above equation $\vec A$ is the vector potential defined implicitly by the equation:
\beq
 \vec B = \vec \nabla \times \vec A.
\label{vecpot}
\enq
The other topological constant is the magnetic cross helicity:
\beq
{\cal H}_C \equiv \int \vec B \cdot \vec v d^3 x,
\label{chel}
\enq
characterizing the degree of cross knottiness of the magnetic field and velocity
lines.

\subsection {Representation in terms of the magnetohydrodynamic potentials}

Let us write the topological constants given in \ern{maghel} and \ern{chel} in terms
of the magnetohydrodynamic potentials $\alpha, \beta, \chi, \eta, \nu, \rho$ introduced
in previous sections.

First let us combine \ern{Bsakurai} with \ern{vecpot} to obtain the equation:
\beq
\vec \nabla \times (\vec A- \chi \vec \nabla \eta) =0,
\label{vecpot2}
\enq
this leads immediately to the result:
\beq
\vec A =  \chi \vec \nabla \eta + \vec \nabla \zeta,
\label{vecpot3}
\enq
in which $\zeta$ is some function. Let us now calculate the scalar product $\vec B \cdot \vec A$:
\beq
\vec B \cdot \vec A =  (\vec \nabla \chi \times \vec \nabla \eta) \cdot \vec \nabla \zeta.
\label{vecpot4}
\enq
However, since we have the local vector basis: $(\vec \nabla \chi, \vec \nabla \eta,\vec \nabla \mu)$
we can write $\vec \nabla \zeta$ as:
\beq
\vec \nabla \zeta = \frac{\partial \zeta}{\partial \chi} \vec \nabla \chi +
\frac{\partial \zeta}{\partial \mu} \vec \nabla \mu + \frac{\partial \zeta}{\partial \eta} \vec \nabla \eta.
\label{zetadiv}
\enq
Hence we can write:
\beq
\vec B \cdot \vec A = \frac{\partial \zeta}{\partial \mu}
(\vec \nabla \chi \times \vec \nabla \eta) \cdot  \vec \nabla \mu =
\frac{\partial \zeta}{\partial \mu} \frac{\partial(\chi,\eta,\mu)}{\partial(x,y,z)} .
\label{vecpot5}
\enq
Now we can insert \er{vecpot5} into \er{maghel} to obtain the expression:
\beq
{\cal H}_M = \int \frac{\partial \zeta}{\partial \mu} d \mu  d\chi d\eta.
\label{maghel2}
\enq
We can think about the magnetohydrodynamic domain as composed of thin closed tubes of
magnetic lines each labelled by $(\chi, \eta)$. Performing the integration along such a thin tube
in the metage direction results in:
\beq
\oint_{\chi, \eta} \frac{\partial \zeta}{\partial \mu} d \mu = [\zeta]_{\chi, \eta},
\label{zetacut}
\enq
in which $[\zeta]_{\chi, \eta}$ is the discontinuity of the function $\zeta$ along
its cut. Thus a thin tube of magnetic lines in which $\zeta$ is single valued does not
contribute to the magnetic helicity integral. Inserting \ern{zetacut} into \ern{maghel2}
will result in:
\beq
{\cal H}_M = \int [\zeta]_{\chi, \eta}  d\chi d\eta= \int [\zeta] d\Phi,
\label{maghel3}
\enq
in which we have used \ern{phichieta}. Hence:
\beq
[\zeta] = \frac{d{\cal H}_M}{d\Phi},
\enq
the discontinuity of $\zeta$ is thus the density of magnetic helicity per unit of magnetic flux in a tube.
We deduce that the Sakurai representation does not entail zero magnetic helicity, rather it is
perfectly consistent with non zero magnetic helicity as was demonstrated above and in agreement to
the claims made by Frenkel, Levich \& Stilman \cite{FLS}. Notice however, that the topological structure
of the magnetohydrodynamic flow constrain the gauge freedom which is usually attributed to
vector potential $\vec A$ and limits it to single valued functions. Moreover, while the choice of
$\vec A$ is arbitrary since one can add to $\vec A$ an arbitrary gradient of a single valued function
 which may lead to different choices of $\zeta$ the discontinuity value $[\zeta]$ is not arbitrary and
 has a physical meaning given above.

Let us now introduce the velocity expression given in \ern{vform} and calculate
the scalar product of $\vec B$ and $\vec v$, using the same arguments as in the previous
paragraph will lead to the expression:
\beq
\vec v \cdot \vec B = \frac{\partial \nu}{\partial \mu}
(\vec \nabla \chi \times \vec \nabla \eta) \cdot  \vec \nabla \mu
= \frac{\partial \nu}{\partial \mu} \frac{\partial(\chi,\eta,\mu)}{\partial(x,y,z)}.
\label{vdotB}
\enq
Inserting \ern{vdotB} into \ern{chel} will result in:
\beq
{\cal H}_C = \int \frac{\partial \nu}{\partial \mu} d \mu  d\chi d\eta.
\label{chel2}
\enq
We can think about the magnetohydrodynamic domain as composed of thin closed tubes of
magnetic lines each labelled by $(\chi, \eta)$. Performing the integration along such a thin tube
in the metage direction results in:
\beq
\oint_{\chi, \eta} \frac{\partial \nu}{\partial \mu} d \mu = [\nu]_{\chi, \eta},
\label{nucut}
\enq
in which $[\nu]_{\chi, \eta}$ is the discontinuity of the function $\nu$ along
its cut. Thus a thin tube of magnetic lines in which $\nu$ is single valued does not
contribute to the cross helicity integral. Inserting \ern{nucut} into \ern{chel2}
will result in:
\beq
{\cal H}_C= \int [\nu]_{\chi, \eta}  d\chi d\eta= \int [\nu] d\Phi,
\label{chel3}
\enq
in which we have used \ern{phichieta}. Hence:
\beq
[\nu] = \frac{d{\cal H}_C}{d\Phi},
\label{cheldensity}
\enq
the discontinuity of $\nu$ is thus the density of cross helicity per unit of magnetic flux.
We deduce that a flow with null cross helicity will have a single valued $\nu$ function
alternatively, a non single valued $\nu$ will entail a non zero cross helicity.
Furthermore, from \ern{nueq} it is obvious that:
\beq
\frac{d [\nu]}{d t} = 0.
\label{nucuteq}
\enq
We conclude that not only is the magnetic cross helicity conserved as an integral
quantity of the entire magnetohydrodynamic domain but also the (local) density of cross helicity per
unit of magnetic flux is a conserved quantity as well.

In the following sub section we give a simple example which
will demonstrate some of the general assertions of this paragraph.

\subsection {A Helical Stratified Magnetic Field}

Consider a magnetohydrodynamic flow of uniform density $\rho$. Furthermore assume that
the flow contains a helical stratified magnetic field:
\beq
\vec B = \left\{%
\begin{array}{ll}
   2 B_{\bot}(1-\frac{R}{a})\hat{\phi} + B_{z0}\hat{z} & R<a \\
    0 & R>a \\
\end{array}%
\right.
\label{Bstra}
\enq
in which $R,\phi,z$ are the standard cylindrical coordinates, $\hat{R},\hat{\phi},\hat{z}$ are the
corresponding unit vectors and $B_{z0},B_{\bot}$ are constants. The magnetic field is contained
in a cylinder of Radius $a$ and is independent of $z$. A possible choice of the vector potential $\vec A$ is:
\beq
\vec A = \left\{%
\begin{array}{ll}
   B_{z0} x \hat{y} + B_{\bot} a (1-\frac{R}{a})^2 \hat{z}, & R<a \\
    0 & R>a \\
\end{array}%
\right.
\label{Astra}
\enq
in which $\hat{y}$ is a unit vector in the $y$ direction. Let us calculate the magnetic helicity
of the field using \er{maghel}. In order to obtain a finite magnetic helicity we assume that the field is contained
between the planes $z=0$ and $z=1$ furthermore we assume that the planes $z=0$ and $z=1$ can be identified
such that the magnetic field lines are closed. Thus the domain becomes a topological torus. Inserting
\ern{Bstra} and \ern{Astra} into \ern{maghel} will result in:
\beq
{\cal H}_M \equiv \int \vec B \cdot \vec A d^3 x = \frac{\pi}{3} a^3 B_{z0} B_{\bot}
\label{maghel2}
\enq
First let us calculate to load using \ern{Load} (we assume that $R<a$ in the following calculations)  we obtain:
\beq
\lambda = \rho \frac{4 \pi B_{\bot}(1-\frac{R}{a}) R + B_{z0}}{B_{z0}^2+4 B_{\bot}^2 (1-\frac{R}{a})^2 } = \lambda(R) ,
\enq
hence the load surfaces are cylinders. The $\chi$ function can now be calculated according to
\ern{chidef} to yield the value:
\beq
\chi =  \frac{1}{2} B_{z0} R^2.
\label{chistar}
\enq
Solving \ern{Bsakurai3} for $\eta$ we obtain the following non unique solution:
\beq
\eta =  \frac{2 B_{\bot}}{B_{z0}} (1-\frac{R}{a}) \frac{z}{a} + \phi .
\label{etastar}
\enq
Substituting \ern{chistar}, \ern{etastar} and \ern{Astra} into \ern{vecpot3} we can solve for $\zeta$
and obtain:
\beq
\zeta =B_{\bot} z (a-R) + \frac{1}{2} B_{z0} x y .
\label{Astar2}
\enq
Since we have identified the $z=0$ and $z=1$ planes the $z$ coordinate is not single valued
and therefore $\zeta$ is a non single valued function which has a discontinuity value:
\beq
[\zeta] =  B_{\bot} (a-R).
\label{etastardis}
\enq
Thus we can calculate the magnetic helicity using \ern{maghel3} and obtain:
\beq
{\cal H}_M =  \int [\zeta] d\Phi  = \frac{\pi}{3} a^3 B_{z0} B_{\bot},
\label{maghel3star}
\enq
which coincides with the result of \ern{maghel2}.

\subsection {Cross Helicity Conservation via the Noether Theorem}

The conservation of helicity $ \int \vec v \cdot \vec \omega  d^3x $ in
ideal (non-magnetic) barotropic fluid when certain conditions are satisfied in
particular, when $ \vec \omega  \cdot \vec n = 0 $ on the (Lagrangian) surface bounding
the volume of integration was discovered by Moffat \cite{Moff2}.
Moreau \cite{Moreau} has discussed the conservation of helicity from
the group theoretical point of view. In his paper he used
an enlarged  Arnold  symmetry group \cite{Arnold0} of fluid element labelling to
generate the conservation of helicity. Yahalom \cite{Yahalomhel,Asher} has shown that
 the symmetry group generating conservation of helicity becomes a very simple one parameter translation
group in the space of labels (alpha space) when represented by Lynden-Bell and
Katz \cite{LynanKatz} labelling. We will now show that in the case of magnetohydrodynamics the same
one parameter translation group will generate the magnetic cross helicity via the Noether Theorem.

Let us denote the initial position of a fluid element by $ (x^k_0) $,
 by mass conservation:
\beq
\rho(x^k)d^3x = \rho(x^k_0) d^3x_0 = \rho(x^k_0){\partial (x^1_0, x^2_0, x^3_0)
\over \partial (\alpha^1,\alpha^2,\alpha^3)}d^3 \alpha.
\enq
Since the initial position of a fluid element can not depend on time it must depend on
 the label only, and therefore by an appropriate choice of the $ \alpha 's $ we obtain:
\beq
\rho(x^k)d^3x = d^3\alpha , \qquad \rho={\partial (\alpha^1, \alpha^2, \alpha^3)
 \over \partial (x^1, x^2, x^3)}.
\label{rhoalph}
\enq
Where we assume that the above expressions of course exist.
Let us look at the action $A$  defined in \ern{Lagaction}.

From the discussion following \ern{delLagaction3} we know that if the $\vec \xi$ variations
disappear at times ${t_0},{t_1}$ than A is
extremal only if  Euler's equations are satisfied and the boundary term disappears. If on the other hand we
make a symmetry displacement i.e. a displacement that makes $\delta A$ vanish
and assume that Euler's equations are satisfied and the boundary term disappears, we obtain that:
\beq
\int_V \vec v \cdot \vec \xi  d^3\alpha = const.
\label{Noether}
\enq
this is Noether's theorem in it's fluid mechanical form.

The $\alpha's$ so chosen as to satisfy the \ern{rhoalph}
are not unique in fact one can always choose another set of variables say
$\tilde \alpha's$ such that:
\beq
 {\partial (\tilde\alpha^1,\tilde\alpha^2,\tilde\alpha^3) \over
 \partial (\alpha^1,\alpha^2,\alpha^3)} = 1.
 \label{alphasym}
  \enq
It is quite clear that if the domain of integration is not modified any new set of
 $\alpha's$ satisfying \ern{alphasym} can be chosen without effecting the value of the Lagrangian $L$.
This is nothing but Arnolds \cite{Arnold0} alpha space symmetry group under which $L$ is invariant
(see also Katz \& Lynden-Bell \cite{KLB}). For some flows the domain of  integration can be modified
 without effecting $L$, in that case we have additional elements in our symmetry group.
 If we make only small changes $\delta \alpha$ than we can define the group as follows:
 \beq
{\partial \delta \alpha_k \over \partial \alpha_k} = 0, \qquad
\delta\vec\alpha\cdot \vec n\bigg|_{surface} = 0,
 \label{smallalphasym}
\enq
where $\vec n$ is a unit vector orthogonal to the surface of the alpha space volume
 which we integrate over. The restriction $\delta\vec\alpha\cdot \vec n\bigg|_{surface} = 0$ is only
 needed when the infinitesimal
 transformation changes the domain of integration in such a way as to modify $L$.
 In this paper we are interested in the subgroup of translation, i.e.:
 \beq
\delta \alpha_k =a_k, \qquad a_k =const.
\label{alphatrans}
\enq
 This subgroup of course  does not satisfy $\delta\vec\alpha\cdot \vec n\bigg|_{surface} = 0$ unless
 at least few of the $\alpha's$ are cyclic or $L$ is not effected by the modification
 of domain.

 In section \ref{inverse} we have defined the three  following  parameters: the magnetic load $\lambda$,
 the magnetic metage $\mu$ and $\eta$.
 Notice that since the magnetic lines are closed $\mu$ is an angular variable and we can translate it
 with out changing $L$. Choosing $\alpha^k = \chi, \eta, \mu$, and inserting those variables
 into \ern{rhoalph} we re-derive \ern{metagecon2}.

 The appropriate $\vec \xi $ symmetry displacement associated with the infinitesimal change
in $\alpha_k$ is given by \ern{xialdef}. For a metage displacement $\vec \xi $ takes the form:
\beq
 \vec \xi = -{\partial \vec r \over \partial \mu}\delta \mu = -\delta \mu \frac{\vec B}{\rho}.
\label{ximu}
 \enq
Inserting this expression into the boundary term in \ern{delLagaction3} will result in:
\beq
\delta A_B =
\int dt  \oint d \vec S \cdot [\vec B (\frac{1}{2} \vec v^2 - w (\rho))
-  \vec v (\vec v \cdot \vec B) ] = 0,
\label{dAB}
\enq
which is indeed Moffat's condition for magnetic cross helicity conservation \cite{Moff2}
 as expected. Inserting \ern{ximu} into \ern{Noether}  we obtain the conservation law:
\beq
\int_V \vec v \cdot {\partial \vec r \over \partial \mu} d^3\alpha =
\int_{V} \vec v \cdot \vec B d^3x = {\cal H}_C.
\label{crhelcon}
\enq
Thus we conclude that the alpha translation group in the direction of $\mu$ generates conservation of helicity.
[ One could of course introduce the symmetry displacement $\vec \xi = \epsilon {\vec B
\over \rho}$, however, in this case one should show that the above
displacement is a symmetry group displacement
which is not obvious if we do not take into account Arnold's
group and Lynden-Bell \& Katz labelling. Moreover in coordinate space the symmetry group appears arbitrary
complex depending on the flow considered as opposed to its apparent simplicity in alpha space.]

\section {Conclusion}

In this paper we have reviewed variational principles for barotropic magnetohydrodynamics given by
previous authors both in Lagrangian and Eulerian form. Furthermore,
we introduced our own Eulerian variational principles from which all the relevant
 equations of barotropic magnetohydrodynamics can be derived and which are in some sense
 simpler than those considered earlier.
 The variational principle was given in terms of six independent functions
for non-stationary flows and three independent functions for stationary flows. This is less
then the seven variables which appear in the standard equations of magnetohydrodynamics which
are the magnetic field $\vec B$ the velocity field $\vec v$ and the density $\rho$.

The equations in the non-stationary case appear have some resemblance to the \eqs deduced in a
previous paper by Frenkel, Levich and Stilman \cite{FLS}. However, in this previous
work the equations were deduced from a postulated Hamiltonian. In the current work
we show how this Hamiltonian can be obtained from our simplified Lagrangian using the canonical Hamiltonian
formalism.

The appearance of a non zero magnetic helicity and cross helicity, is connected with the
fact that some of the functions which we defined are non-single valued. This was elaborated
to some extent in the final section of this paper and was connected to the properties of the
functions $\zeta,\nu$. We have also shown that the density of cross helicity per unit of magnetic flux
is also a conserved quantity and is equal to the discontinuity of $\nu$.
Furthermore, we have shown that the conservation of cross helicity
can be deduced using the Noether theorem from the symmetry group of magnetic metage translations.

The problem of stability analysis and the description of
numerical schemes using the described variational principles exceed the scope of this paper.
We suspect that for achieving this we will need to add additional
constants of motion constraints to the action as was done by \cite{Arnold1,Arnold2}
see also \cite{YahalomKatz}, hopefully this will be discussed in a future paper.
\\
\\
\noindent
{\Large \bf Acknowledgement}
\\
\\
\noindent
The authors would like to thank Prof. H. K. Moffatt for a useful discussion.

\end {document}